\documentclass[12pt]{article}

\usepackage{amssymb}
\usepackage{amsmath}
\usepackage{cite}
\usepackage{epsf}

\newcommand{\sfrac}[2]{{\textstyle\frac{#1}{#2}}}

\renewcommand{\thefootnote}{\arabic{footnote}}
\newcommand\ZZ{\hbox{\zfont Z\kern-.4emZ}}
\font\zfont = cmss10 

\renewcommand{\sfrac}[2]{{\textstyle\frac{#1}{#2}}}

\def\GNew{\mathcal{G}_N}
\def\sinn{\mbox{sinn}}

\begin{document}

\begin{titlepage}
\begin{flushright}
Saclay T01/126\\
{\tt hep-ph/0111335} \\
\end{flushright}

\vskip.5cm
\begin{center}
{\huge{\bf Supernovae as a Probe of}}
\vskip.2cm
{\huge{\bf  Particle Physics and Cosmology}}
\vskip.2cm
\end{center}
\vskip0.2cm

\begin{center}
{\sc Joshua Erlich}$^{a}$ and
{\sc Christophe Grojean}$^{b}$ \\

\end{center}
\vskip 10pt

\begin{center}
$^{a}$ {\it Theory Division T--8,
Los Alamos National Laboratory, Los Alamos, NM 87545, USA} \\
\vspace*{0.1cm}
$^{b}$ {\it Service de Physique Th\'eorique, CEA--Saclay,
F--91191 Gif--sur--Yvette, France} \\
\vspace*{0.1cm}
{\tt  erlich@lanl.gov, grojean@spht.saclay.cea.fr}
\end{center}

\vglue 0.3truecm

\begin{abstract}
\vskip 3pt \noindent
It has very recently been demonstrated by Cs\'aki, Kaloper and Terning
(CKT)
that the faintness of supernovae at high redshift can be accommodated
by mixing of a light axion with the photon in the presence of an
intergalactic magnetic field, as opposed to the usual explanation of
an accelerating universe by a dark energy component.
In this paper we analyze further aspects of the CKT mechanism
and its generalizations.  
The CKT mechanism also passes various cosmological
constraints from the fluctuations of the CMB and the formation
of structure at large scales, without requiring an accelerating phase in
the expansion of the Universe.
We investigate the statistical significance
of current supernovae data for
pinning down the different components of the cosmological 
energy-momentum tensor
and for probing physics beyond the standard
model.
\end{abstract}

\end{titlepage}

\newpage

\renewcommand{\thefootnote}{(\arabic{footnote})}

\section{Introduction}
\label{sec:intro}
\setcounter{equation}{0}
\setcounter{footnote}{0}

The dynamics of the Universe in the standard cosmological model 
is completely determined by the present
values~\footnote{A subsript $0$ will always denote
the value taken today by any quantity.}
of the Hubble parameter, $H_0$, and the ratios $\Omega_{0 i}$ between
the energy density of the different ``matter" components and the critical 
density.
The task assigned to cosmologists by Sandage is the precise determination
of these numbers. Unfortunately, they are not directly measurable and their
determination relies on the measurement of quantities indirectly
dependent on them.
Several bounds on the $\Omega_{0 i}$ follow from the study of the anisotropies
of the cosmic microwave background radiation (CMB), the formation of
large scale structure, the age of the Universe, {\it etc.} The Hubble
diagrams (luminosity distances vs. redshifts) of standard candles have
traditionally
been a useful tool for determining constraints on
some of the cosmological parameters.  One of
the most
astonishing physical results of the past few years \cite{HZT,SCP} is that not 
only is 
the Hubble diagram of high redshift supernovae (SNe) compatible
with previous independent constraints on the $\Omega_{0 i}$, but it is also
complementary to them and in principle allows a complete determination
($H_0\sim 66\, \mathrm{\, km.s}^{-1}.\mathrm{Mpc}^{-1}$,
$\Omega_{0 m}\sim.3$ and $\Omega_{0 \Lambda}\sim.7$). In particular this result
reveals the presence of a so-called ``dark energy'' component accounting
for the dimming of high-$z$ SNe as a result of a current period of
acceleration in the expansion of the Universe. However, both the existence
as well as the size of this Dark Energy component jeopardize
our current understanding of the fundamental laws of particle 
physics~\cite{ccQFT}
and string theory/quantum gravity~\cite{ccString}.

Very recently Cs\'aki, Kaloper and Terning (CKT)\cite{CKT} have proposed an
interesting alternative explanation of combined Type Ia SNe data from the
Cal\'an-Tololo \cite{C-T} survey, the Supernova Cosmology Project (SCP)
\cite{SCP}, and the High-$z$ Supernova Search Team (HZT) \cite{HZT} without
requiring a period of accelerating expansion.
By assuming the existence of a light axion $\phi$ 
($m_\phi\sim 10^{-16}$ eV$/c^2$) and an
intergalactic magnetic field of amplitude $|\mathbf{B}|\sim 10^{-13}$ Tesla,
the authors of \cite{CKT} found that the magnitude-redshift curve corresponding
to the best fit flat universe with a cosmological constant~\cite{SCP}
($\Omega_{0 m}\sim 0.28, \Omega_{0 \Lambda}\sim 0.72$) can be closely reproduced with
cosmic strings replacing the cosmological constant.  The axion interacts with photons
via a term in the effective Lagrangian of the form,
${\cal L}_{int}=\frac{1}{M_L\, c^2}\, \phi \,\mathbf{E}\cdot\mathbf{B}$,
where $\phi$ is the axion field and $M_L$ is the axion coupling mass scale, 
which
CKT fit to $M_L\sim 4\cdot 10^{11}$ GeV$/c^2$.  As a result of this coupling,
photons and axions will oscillate in a background magnetic field via a
Primakoff effect.  Assuming
a randomly oriented magnetic field with a coherence length of the order of
1 Mpc, almost one third of the highest redshift photons would decay by the
time they reach us.

In this letter we study the statistical significance of the CKT results, and
discuss possible generalizations of the CKT mechanism.
In order to understand the model dependence of generalized photon loss
mechanisms we use the SNe data to identify the allowed regions in the
parameter space of such models, obtaining fits at least as good as 
the cosmological constant accelerating universe. 
The anisotropy of the CMB still requires a Dark Energy component to
the energy-momentum tensor.  The large scale structure of the Universe
then imposes constraints
on the equation of state of the invisible Dark Energy. 
The difference from previous analyses is that the SNe data now allows for a 
significantly larger range of the equation of state parameter $w_X$ 
(in $p=w_X\rho$) for the Dark Energy component. It is interesting that the
preferred region for $w_X$ remains in the range $w_X<0$, in agreement with 
requirements of large scale structure.
The key point is that instead of indicating
an accelerating expansion of the Universe, the dimming of
SNe at high redshift would reveal the presence of a light axion 
which mixes with photons in the intergalactic medium.

Because this subject unites
several branches of theory and experiment which some readers may not be
familiar with, in Section~\ref{sec:bounds} we
briefly review the relevant experiments and current bounds on the relevant
cosmological and particle physics parameters.
In Section~\ref{sec:theory} we review the theoretical
basis of the CKT photon loss mechanism and its generalizations. This 
mechanism
is intrinsically quantum mechanical, as opposed to the essentially classical
models of photon absorption by dust~\cite{dust}.
We then describe in Section~\ref{sec:cosmography} the constraints
placed on the cosmological parameters by the Hubble diagrams of Type Ia SNe and
how they are modified in the presence of a photon loss mechanism.

In Section~\ref{sec:fits} we perform a statistical analysis of the data and
our fits, and demonstrate that the CKT model with a best fit photon decay
parameter (with $\Omega_{0 m}=.3,\,\Omega_{0 X}=.7, w_X=-1/3$) is as good as the
$\Omega_{0 m}=.28,\,\Omega_{0 \Lambda}=.72$ cosmological constant accelerating 
universe scenario.
We present some exclusion regions of the
parameter space, but there is a large range
of allowed parameters for a generalized CKT mechanism to accommodate the data.

We conclude with proposals for future experiments to
better pin down the Sandage numbers and ferret out the nature
of the Dark Energy, and perhaps reveal the existence
of an axion coupled to the particles of the Standard Model or other new 
physics beyond the standard model.

\section{Experiments and bounds}
\label{sec:bounds}
In this section we summarize experimental bounds on and estimates of
the relevant cosmological,
astrophysical and particle physics parameters.

\subsection{Cosmological Constraints}

Although the energy density ratios $\Omega_{0i}$
are not directly measurable, they affect the time evolution
of the scale factor of the Universe which in turn controls at least three 
types
of measurable quantities: (i) the growth of density perturbations 
responsible for the formation
of structure at large scales; (ii) the age of the Universe; (iii) the 
propagation of light between its emission by some known astrophysical
sources and its detection today.

The sequence of structure formation starting from smaller fluctuations and 
ending with clusters of galaxies is well 
described by cold dark matter models, and the structure  
observed today could well have
evolved from the density perturbations revealed by the CMB anisotropy at
$z \sim 1100$. The presence
of Dark Energy should not inhibit the growth of density perturbations, which 
implies that the Dark Energy 
should  have started to dominate only recently.  This requires 
an upper bound on $w_X=p_X/\rho_X$ at least as strong as
$w_X < w_m =0$~\cite{TurnerWhite, PTW}.

The fluctuations in the CMB can also be used to predict the distribution of matter, baryon 
and Dark Energy in the Universe. The position of the first peak in the power spectrum as a
function of multipole moment depends on the total density
($\Omega_{0 m}+\Omega_{0 X}$) and also less sensitively
on $\Omega_{0 m}$ and $\Omega_\mathrm{0 baryon}$.
The latest determination by the BOOMERANG experiment is about
$\Omega_0=1.0\pm 0.1$ and $\Omega_{0 m} \sim .33\pm .03$ (and only a
small fraction of baryons)~\cite{BOOMERANG}.
The flat universe is heavily favored by this observation, as any deviation
from flatness tends to grow quickly in Big Bang cosmology. This observational result
as well as a nearly scale-invariant spectrum of primeval density
perturbations
is in accordance with the predictions of inflationary models.
The estimate of $\Omega_{0 m}$ seems also in agreement, at least for its order of magnitude,
with the measurements of mass-to-light ratios in clusters. Even if the CMB
cannot directly teach us about the Dark Energy
(the CMB is an image of the Universe at $z\sim 1000$ where the Dark Energy
was completely washed out), it still points out a missing component of
the energy-momentum tensor.

Some other cosmological constraints, for example from the age of the
Universe or gravitational lensing,
are less predictive due to high uncertainties and limited data.

Finally, we can determine some cosmological constraints from the Hubble
diagrams of some ``standard candle'' sources: if we know the
absolute luminosity of the source, by measuring the flux we receive today,
we can deduce the distance travelled by the light, and this distance
depends on the $\Omega_{0 i}$ ratios. At very low redshift ($z < .1$), this
dependence is universal while at large redshift ($z \sim 5$) all
the components with $w_i<0$ have been washed out by the matter
energy density. The intermediate interval
is thus the best suited to bring some information on
the Dark Energy. Among the astrophysical objects observed in
this redshift interval are the Type Ia SNe. And by chance,
Type Ia SNe have been found to be decent standard candles by means
of matching luminosity vs.~time curves to the phenomenological Phillips
curve \cite{Phillips}.  As we will explain in detail later, the information
on the cosmological parameters inferred from the Hubble diagrams
relies on the nature of the propagation of light. Assuming a free propagation
of the photons in the cosmological background, and assuming a flat universe,
the best fit of the SNe data gives~\cite{SCP}
$\Omega_{0 m}=0.28\pm0.09$.
This is in good agreement with several other
determinations, which is perhaps why the relatively poor current statistics
in high-$z$ SNe data is sometimes overlooked.
The aim of this work is to study how this conclusion is modified
when the photons are  no longer freely propagating but rather interact
with an axion in such a way that some photons are lost
between their emission by the SNe and their detection today.
The following section will review the CKT photon loss mechanism
and we will analyze the SNe data taking into account
such a photon loss mechanism in Section~\ref{sec:fits}.

\subsection{Astro-particle Constraints}

The axion is a proposed ultralight pseudoscalar  field that would 
couple weakly with ordinary stable matter.  
The relevant coupling for the present 
study is that between axions and photons.
The coupling is through a term in the Lagrangian of the form,
${\cal L}\supset \frac{1}{M_L \,c^2}\, \phi \,\mathbf{E}\cdot\mathbf{B}$, 
where $\phi$ is the
axion and $M_L$ is the axion coupling mass scale.  
Several searches for axions
have turned up negative, placing constraints on the axion mass and
couplings.  In particular, a study of conversion of solar axions to 
X-rays in a strong magnetic field give 
$M_L > 1.7  \times 10^{9 }$~GeV$/c^2$~\cite{Moriyama}.
Based on mixing between axions and photons in an external
magnetic field, precisely the effect needed for the CKT mechanism, 
studies of globular-cluster stars provide a rough limit 
$M_L > 1.7 \times 10^{10}$~GeV$/c^2$~\cite{Raffelt}.  For more complete
reviews, see \cite{axion-reviews}.  Bounds on the axion
mass are closely tied to the axion coupling, but the absence of intergalactic
line emissions and other experiments place 
a rough bound of $m_\phi <10^{-3}$~eV$/c^2$~\cite{axion-mass}. The CKT axion
of mass $m_\phi \sim 10^{-16}$~eV$/c^2$ and coupling of 
$M_L  \sim 4 \cdot 10^{11}$~GeV$/c^2$
is within current bounds.  Some further studies and bounds can be found in 
\cite{Toldra}.

The  mixing
between the axion and the photon requires the existence
of the magnetic field.
Magnetic fields appear to permeate the observable universe 
over a large range of
scales, from relatively strong fields 
in the interstellar galactic medium, to weaker fields in the outer
envelopes of X-ray clusters.  The intergalactic magnetic field can be
deduced by observations of radiation from relativistic electrons
around X-ray clusters.  These observations suggest an intergalactic magnetic
field of around $B\sim 10^{-13}$ Tesla.  (See~\cite{Furlanetto-Loeb,Gnedin} for
recent summaries.)  The magnitude of the
intergalactic magnetic field can also be estimated by other means, 
including studies of radio 
emission of distant quasars which, assuming a coherence length for the 
magnetic field of about 1 Mpc, yields roughly the same 
estimate~\cite{Kronberg}.
The source of these magnetic fields is not well understood, but there
have been several suggestions made, from inhomogeneous cosmological lepton
number \cite{Dolgov-Grasso} to expanding quasar 
outflows~\cite{Furlanetto-Loeb}.  
Based on these results, CKT assumed a bulk magnetic field of
$10^{-13}$ Tesla, and assumed a coherence length of 1 Mpc.
The distribution and variation of the magnetic fields in the 
intergalactic medium is still not well understood.
For a recent review of the relevant experiments
and the significance of magnetic fields in the early universe, and for a
more complete
list of references, see~\cite{Grasso-Rubinstein}.

\section{CKT photon loss mechanism}
\label{sec:theory}

Following the pioneering works~\cite{PhotonMixing} of Sikivie, 
the CKT model considers an axion--photon
coupling in the inter-galactic magnetic field \footnote{
Throughout the paper, we will use electromagnetic units
such that both the electric and magnetic fields
have the dimension of squared energy while the axion scalar field
has the dimension of energy. This means that
the magnetic field in SI units is given by
$B_{SI}=|\mathbf{B}|/\sqrt{4\pi \epsilon_0 \hbar^3 c^5}$.
The speed of light will be denoted by
$c$, Newton's constant by $\GNew$, and the vacuum permittivity by
$\epsilon_0$.}
\begin{equation}
\mathcal{L}_{int}=\frac{1}{M_L \, c^2}\, \phi \,\mathbf{E}\cdot\mathbf{B}
\end{equation}
where the scale $M_L$ characterizes the strength of the axion--photon interactions.
In a region where the magnetic field is approximately constant, the
polarization
of the photon whose electric field is parallel to the magnetic field mixes
with the axion
to form an oscillating system, much like that of the massive neutrino
system. The mass eigenstates in the presence of the background magnetic field
are not the same as the mass eigenstates in vacuum, and thus the photons
oscillate into
axions and {\it vice et versa.} A quantum mechanical computation
gives the probability that a photon has not yet oscillated into an axion
over a distance $l$ in a uniform magnetic field as \cite{CKT},
\begin{equation}
P_{\gamma \to \gamma} (l) =
1 - \frac{4\, \mu^2 \epsilon^2}{m_\phi^4 \,c^8 + 4\, \mu^2 \epsilon^2}
\sin^2 \left(
(\sqrt{\epsilon^2-\lambda_+} - \sqrt{\epsilon^2-\lambda_-}) l/(2 \hbar c)
\right) ,
\end{equation}
where $m_\phi$ is the mass of the axion, $\epsilon$ is the energy of the photon,
$\mu=|\mathbf{B}|/(M_L\, c^2)$ and
$\lambda_{\pm}=(m_\phi^2\, c^4 \pm \sqrt{m_\phi^4 \, c^8+ 4 \mu^2 \epsilon^2})/2$
are the squared-mass eigenvalues.  Note that despite the fact
that $\lambda_-<0$, that mode has positive energy when the energy
$\epsilon$ is taken into account.

If $\mu\ll m_\phi c^2$ then there are two distinct energy regimes:
\begin{itemize}
\item High energy photons ($\epsilon \gg m_\phi^2 c^4 /\mu$),
for which the oscillation is maximal and achromatic:
\begin{equation}
P_{\gamma \to \gamma} (l) = 1 - \sin^2 \left( \mu\, l /(2\hbar c) \right)
\end{equation}
\item Low energy photons ($m_\phi c^2 \ll \epsilon \ll m_\phi^2 c^4 /\mu$),
for which
the oscillation is small and energy-dependent:
\begin{equation}
P_{\gamma \to \gamma} (l) =  1 - 4 \, \frac{\mu^2 \epsilon^2}{m_\phi^4\, c^8}
\sin^2  \left( m_\phi^2 c^3 l / (4 \hbar\, \epsilon ) \right)
\end{equation}
\end{itemize}

In the Universe, the intergalactic magnetic field is not uniform but
is assumed by CKT to vary
in domains of size $L_{\mbox{\small dom}} \sim 1 \mbox{Mpc}$,
which is much smaller than the distance to high redshift SNe
($z\sim1$ is equivalent
to distances about $3 \times 10^3$ Mpc). CKT  assume a uniform distribution of
randomly oriented
magnetic fields of constant magnitude over the 1 Mpc domains\footnote{As
mentioned earlier, these
magnetic fields are poorly understood and may be distributed
much more irregularly than is assumed here.  We thank Hui Li for a discussion
of this point.}.  In that case
high energy unpolarized photons (as defined above) approach an equilibrium
population evenly divided between the two photon polarizations and the
axion at large distances.  CKT found that the approach to equilibrium
is through an exponential damping of the number of high energy
photons \cite{CKT}:
\begin{equation}
P_{\gamma \to \gamma} (l) \simeq
\sfrac{2}{3} + \sfrac{1}{3} \, e^{-l/L_{\mathrm{dec}}},
\end{equation}
where the decay length is approximatively given by:
\begin{equation}
L_{\mbox{\small dec}} =  \frac{8}{3} \,
\frac{\hbar^2c^6\, M_L^2}{ L_{\mathrm{dom}}\, |\mathbf{B}|^2}.
\end{equation}

Before studying the implications of this photon loss mechanism
on the SNe data, it is worth mentioning how to evade any unwanted
effects on the CMB anisotropy. While the photons of the CMB  were
emitted at energies of the atomic sizes, when the magnetic field
appears they have already redshifted down to much lower energies about
$10^{-4}$~eV. In this low energy regime, the oscillation
mixing is energy suppressed and requiring  $1-P_{\gamma\to \gamma}$
to be less than the CMB anisotropy ($10^{-5}$) gives a lower
bound on the axion mass around $m_\phi > 10^{-16}$~eV~\cite{CKT}.
\footnote{ It has been noted that the photon loss probability and
the above conclusions may be modified in
the presence of the intergalactic plasma \cite{IGP}, depending on the plasma
density.  It was suggested in \cite{IGP2} that
a small free electron density can in fact weaken the constraints on the axion
parameters.
In any case, we have not included such modifications in this analysis.}

\section{Cosmography: luminosity distance vs. redshift with and without photon loss}
\label{sec:cosmography}

Let us now study the consequences of this photon loss mechanism
on the measurement of luminosity distances.  A nice review
of the cosmological
parameters and their determination for non-specialists
can be found in \cite{Pal}.

A homogeneous and isotropic space, as our Universe appears to be
at large scales, is described by the Robertson--Walker metric
\begin{equation}
ds^2 = - c^2 dt^2 + R_\circ^2 a^2(t) \left(
\frac{dr^2}{1-k r^2} + r^2 d\theta^2 + r^2 \sin^2 \theta d\phi^2 \right)
\end{equation}
where $a(t)$ is the dimensionless scale factor (by convention
$a(\mbox{\small today})=1$),
$R_\circ$ characterizes the absolute
size of the Universe, while with an appropriate rescaling of the coordinates
the curvature parameter $k$ takes values $+1,0,-1$ for closed, flat
and open universes.

For energy-momentum sources made of a superposition of perfect fluids specified by
their energy densities $\rho_i$ and their pressures $p_i$, the Einstein equations reduce
to the Friedmann equation (a dot stands for a derivative with respect to the time coordinate):
\begin{equation}\label{eq:Friedmann}
H^2 = \frac{8 \pi \GNew}{3\, c^2  } \sum_i \rho_i - \frac{k c^2}{a^2 R_\circ^2},\ \ \ \
\mbox{where } \ H=\frac{\dot a }{a}
\end{equation}
together with the conservation equation:
\begin{equation} \label{eq:conservation}
{\dot \rho}_i + 3 (\rho_i+p_i)H =0
\end{equation}
It is common to introduce the ratio of the energy densities to the critical energy density for a flat universe:
\begin{equation}
\Omega_i = \frac{8\pi \GNew \rho_i}{3 \, c^2 \,  H^2}
\ \ \ \mbox{and } \ \ \
\Omega_{\mbox{\small curv.}} = -\frac{k c^2}{a^2 R_\circ^2 H^2}
\end{equation}
Note that $\sum_i \Omega_i +\Omega_{\mbox{\small curv.}}=1$ by
(\ref{eq:Friedmann}).
In terms of the redshift parameter $z=1/a-1$ (by convention, since
$a_0=1$, $z_0=0$)
and assuming
a constant equation of state $p_i=w_i \rho_i$ for each component of the
stress-energy
tensor, the conservation equation (\ref{eq:conservation}) can be integrated
with respect to time, giving a relation between the densities $\rho_i$,
the scale factor $a$ and the equation of state $w_i$.  This gives,
\begin{equation}
\rho_i=\rho_{0i}\,(1+z)^{3(1+w_i)}. \end{equation}
The Friedmann equation (\ref{eq:Friedmann}) can then be rewritten~\footnote{From now we will omit
the subscript $0$ on the energy density ratios $\Omega_i$ evaluating
those quantities only at the present time.}:
\begin{equation}\label{eq:Hsq}
H^2 (z) = H_0^2 (1+z)^2 \left( 1+ \sum_i \Omega_{i} ((1+z)^{1+3w_i} -1) \right)
\ \ \ \mbox{and } \ \ \
1-\sum_i \Omega_{i} = -\frac{k c^2}{R_\circ^2 H_0^2 }
\end{equation}

A crucial test of cosmological models comes from Hubble diagrams for
standard candles by
plotting their luminosity distances versus their redshifts. The luminosity
distance is defined by considering a source
located at the comoving distance $r_s$, which has emitted some light at a
time
$t_s$ with an absolute
luminosity $\mathcal{L}$ (energy per time produced by the source,
normalized to that of a standard candle). This
light is
detected today by a detector located at $r=0$ which measures a flux
$\mathcal{F}$
(energy per time per area) and a redshift $z_s$.
Taking into account the energy loss due to the redshift
of the light and the increase of time intervals between the emission
and the reception,
the conservation of energy ensures that,
\begin{equation}
        \label{eq:L-F}
\mathcal{L} = 4\pi\, r_s^2\, R_\circ^2\, (1+z_s)^2\, \mathcal{F}.
\end{equation}
The luminosity distance between the source and the detector is defined by,
\begin{equation}
\mathcal{L} = 4\pi\, d_L^2\,  \mathcal{F}.
\end{equation}
The quantities characterizing the source are not independent
but can be related to each other  through the ratios $\Omega_{i}$
governing the dynamical evolution of the Universe. This is why
the independent measurement of two SNe observables can tell us
about Sandage's numbers.
The trajectory of light
is given by the null geodesic equation, $ds^2=0$, which when integrated
becomes,
\begin{equation}
\int_{0}^{r_s} \frac{dr}{\sqrt{1-kr^2}} =
\int_{t_s}^{\mathrm{today}} \frac{c\, dt}{R_\circ a(t)} =
\int_{0}^{z_s} \frac{c\, dz}{R_\circ H(z)}
\end{equation}
Using (\ref{eq:Hsq}) this integral equation can be solved to relate the
comoving distance $r_s$
of the source to its redshift\footnote{The
definition of the function `$\sinn$' depends
on the topology of the Universe:
$\sinn x=\sinh x, x, \sin x$ for an open, flat, closed Universe,
respectively.  See for instance~\cite{CPT}.}:
\begin{equation}
        \label{eq:r-z}
r_s (z_s) = \sinn \left( \frac{c}{R_\circ H_0}
\int_{0}^{z_s} \frac{dz}{(1+z) \sqrt{ 1 + \sum_i \Omega_{i} ((1+z)^{1+3w_i}-1)}}
\right)
\end{equation}
This relation allows one
to predict the luminosity distance of the source in terms of its
redshift:
\begin{equation}\label{eq:d_L}
d_L = (1+z_s)\, r_s(z_s) \, R_\circ .
\end{equation}
So far this expression has been derived assuming that once emitted the
light propagates
freely in all spatial directions. The CKT proposal
that photons can disappear into axions in the intergalactic magnetic field
changes the relation (\ref{eq:d_L}) for $d_L(z_s)$.
Imagine indeed that over the distance $r_s R_\circ$ a fraction
$(1-P_{\gamma\to \gamma}(r_s R_\circ))$ of the number of photons
has decayed into axions, so that only a fraction
$P_{\gamma\to \gamma}(r_s R_\circ)$ of the energy emitted by the source will
be detected and the conservation equation (\ref{eq:L-F})
has to be modified accordingly:
\begin{equation}\label{eq:lum}
\mathcal{L} =
4\pi \, r_s^2\,  R_\circ^2\, (1+z_s)^2 \, \mathcal{F} /
P_{\gamma\to \gamma}(r_s R_\circ )
\end{equation}
Then the expression for the luminosity distance as a function
of the redshift becomes:
\begin{equation}
d_L = \frac{(1+z_s)\ r_s(z_s)\, R_\circ}{P^{1/2}_{\gamma\to \gamma} (r_s(z_s)\, R_\circ)}
\end{equation}
where $r_s(z_s)$ is still given by (\ref{eq:r-z}).
Usually, the data are not presented in terms of $d_L(z_s)$ but rather
in terms of the magnitude defined as
\begin{equation}
m=m_0 + 5 \log_{10} (H_0 d_L /c)
\end{equation}
where $m_0$ is the magnitude associated to the absolute
luminosity of the source (after corrections deduced from the fitting to the
Phillips curve, $m_0$ is the common quantity to all SNe and it is fit to the data).

We would like to comment briefly on the effect of photon loss on the measured
flux.  The luminosities of distant SNe fluctuate too much
in general for such SNe to make good standard candles.
However, it has been noticed  that
the absolute luminosities of Type Ia SNe can be determined by
fitting the time evolution of the measured SNe luminosity
to the phenomenological
Phillips curve \cite{Phillips}.  Roughly speaking, the broader the light
curves of a Type Ia SN, the brighter it is intrinsically.
This relation is specified phenomenologically by a stretching factor
relating the width of the light curve to its intrinsic luminosity. Given the
absolute luminosity, together with the measured luminosity, one determines
a normalized flux ${\cal F}$.  The effect of photon loss is to reduce the
measured flux, but it is important that the Phillips curve not change
as a result.  The first point is that the Phillips curve
measures brightness on a logarithmic scale as a fuction of the age of
the SN.  Hence, if an energy-independent fraction of photons
is lost by a particular source, the light
curve will simply be shifted vertically.  This is the effect we have been
discussing.
The only reason for concern is then whether or not the broadness of the
observed light curve should change as a result of photon loss,
as that is how the absolute luminosity
of a Type Ia SN is determined.  The luminosity signal is generally
high enough above background that we believe this will be a negligeable
effect.  A more rigorous analysis should be done, however.

The CKT model described above provides a physical example of
a photon loss mechanism
in which the $P_{\gamma\to \gamma}$ probability is directly related
to particle physics parameters that can therefore be constrained by the SNe
data.

What the SNe data indicate (Figure~\ref{fig:Hubble}) is that
SNe with  $z\sim1$ are dimmer than can be explained by a universe
with only ordinary
matter in it ($w_M=0$), assuming no loss of photons.  The universe undergoes
a period of acceleration if the Dark Energy has an equation of state
$w_X<-1/3$ (the condition for having an accelerating
period of acceleration today is $w_X<-1/(3-3\Omega_m)$ and
if $-1/(3-3\Omega_m)<w_X<-1/3$ the universe will accelerate in the
future).
Suppose we choose to assume that there is no period of
acceleration.
Then we can study the general requirements of a photon loss mechanism to
explain
the SNe data.  We think it is worth studying such generalizations
because the authors of \cite{CKT} make many assumptions regarding
observational
astrophysics which are not well understood.  For example, as mentioned
earlier, the structure of
the intergalactic magnetic field may be significantly different than the
CKT assumption of 1 Mpc domains of constant magnitude.
Furthermore, there may be
other acceptable photon loss mechanisms, either via new particle physics
or via uniform absorption by an unknown intergalactic medium.

For $z< 1$ we require that the photon loss be so as to force a positive
slope for the magnitude-redshift curve.  For example, in the absence of a
photon loss mechanism, the dotted curve in Figure~\ref{fig:Hubble} must be
forced upward.  The SNe magnitude is proportional to
the logarithm of its
luminosity, so by (\ref{eq:lum}) this magnitude is shifted by
the addition of a term proportional to
$\log(P_{\gamma\rightarrow\gamma})$.  If we assume that
for low $z$, $P_{\gamma\rightarrow\gamma}\sim 1-\delta
P_{\gamma\rightarrow\gamma}(z)$ where $\delta P_{\gamma\rightarrow\gamma}(z)
\ll 1$, then we require that $\delta P_{\gamma\rightarrow\gamma}(z)$
have a linear term which can control the slope of the magnitude-redshift
curve.  Otherwise a fit will likely
become more difficult.  The other requirement is that if we want to match
onto
the single Type Ia observation at $z\sim 1.7$~\cite{SN1997}, then the magnitude-redshift
curve must be allowed to slope downward by then.  The easiest way to
accomplish this is to give $\log(P_{\gamma\rightarrow\gamma}(z))$
a small negative slope and positive second derivative near $z\sim 1$.
These requirements are satisfied by the CKT mechanism.  We have
also checked that in a variety of phenomenological photon loss
models (without theoretical basis) for which these
two requirements are satisfied, it is rather easy to fit the
SNe data at least as well as the accelerating universe
by a cosmological constant.

\section{Statistical fits}
\label{sec:fits}
In this section we describe the statistics behind our $\chi^2$
fits for the parameters
in the generalized CKT model.  We have included a total of
18 SNe from the
Cal\'an-Tololo Supernova Survey \cite{C-T} and 36 from the Supernova
Cosmology
Project \cite{SCP}.
Following \cite{SCP} we removed six outlying data points of their 42 total,
corresponding to their Fit C.  Furthermore, in the fits presented here
we have tentatively
dropped the $z\sim 1.7$ point~\cite{SN1997} because of its large error bars compared
to the remaining data.  We have checked that the results are nearly
unchanged when that point is included.

In our fits we have assumed a two component universe
with a fractional density $\Omega_m$ of ordinary matter and $\Omega_X$
of Dark Energy with equation of state $p_X=w_X \rho_X$.  We assume the
CKT model for photon loss parametrized by the photon decay length
$L_{\rm dec}$. Our theoretical models thus involve five parameters:
$\Omega_m$, $\Omega_X$, $w_X$, $m_0$ and $L_{\rm dec}$.
So far all the statistical analyses of the SNe data
have arbitrarly reduced this parameter space by choosing $L_{\rm dec} = \infty$
{\it a priori}. As we will see now, the introduction of this extra parameter
to fit the data changes qualitatively and quantitatively the confidence
levels in the determination of the cosmological parameters.

Because the absolute magnitude, $m_0$, is related to the calibration
of the measurements, we have chosen to systematically fit this variable
to the SNe data (both low and high redshift) with a uniform prior.
Hence the number of parameters in
the fits presented here is always
one greater than the effective theoretical parameter space.
This procedure is slightly
different than that which is sometimes performed, in which the additional
annoyance parameters are integrated over.  Because the distribution in $m_0$
is quite highly peaked, the two procedures are nearly equivalent.  There will
in addition
be a small correction to these fits from the error in the stretching factor
of the Phillips curve (we used the effective magnitudes as presented by the
SCP collaboration, which already involved fitting the stretching factor).
In the various fits presented
in this letter, some further constraints have been placed on the remaining
parameters of the model in order to make use of previous experimental
results and to limit the dimensionality of our confidence plots.

The likelihood of a model to explain experimental data is measured by a
$\chi^2$ together with the number of data points and model parameters, which
are then translated into a confidence level about the best fit:
\begin{equation}
\chi^2 = \sum_{i,j} (m_i^{\rm exp} - m_i^{\rm th}) \sigma_{ij}^{-2}
(m_j^{\rm exp} - m_j^{\rm th}).
\end{equation}
The correlated error matrix $\sigma_{ij}$ can be found
at {\tt http://www-supernova.lbl.gov}  for the SCP data .

\begin{figure}[ht]
\centerline{\epsfxsize=15cm\epsfbox{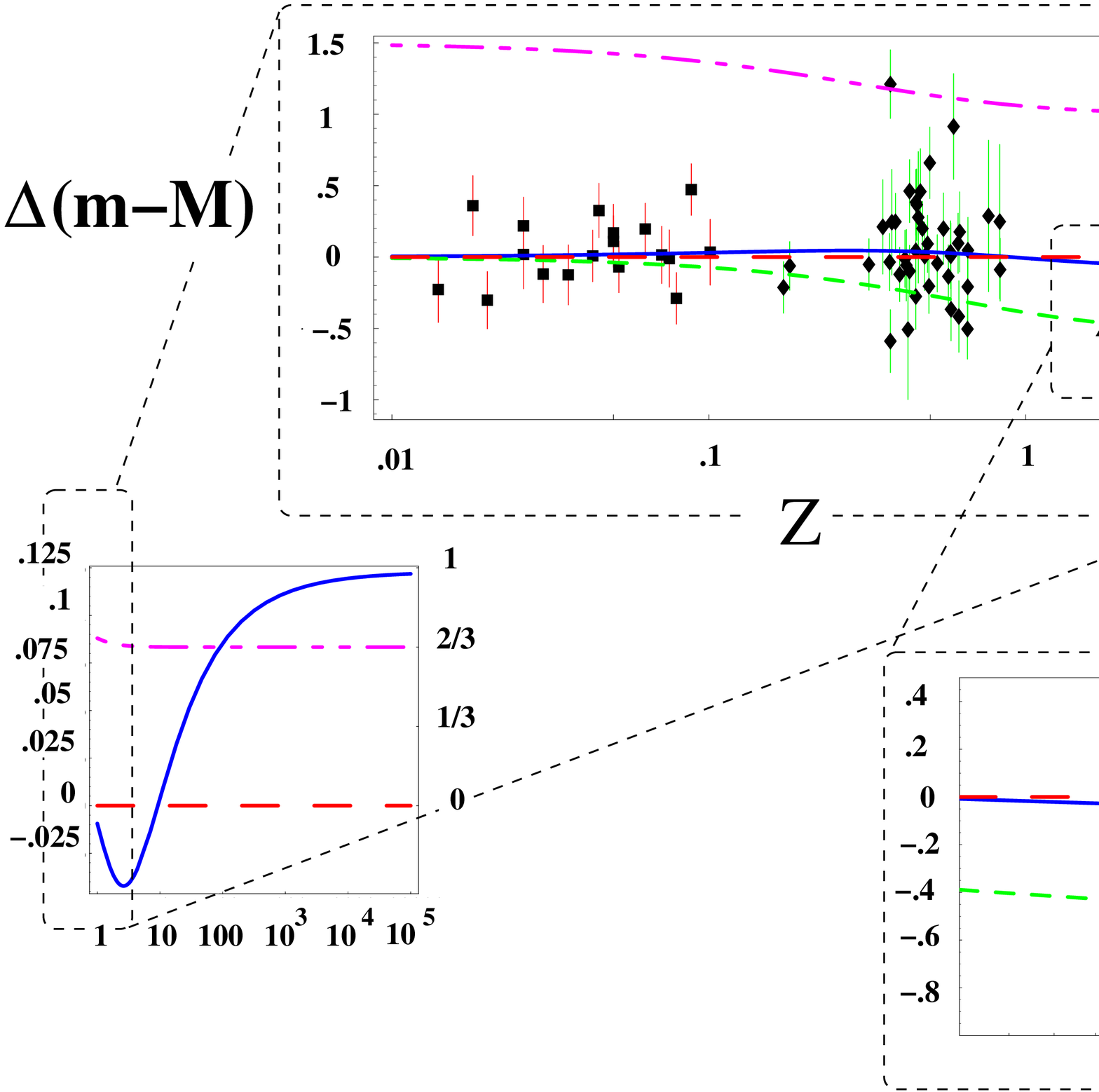}}
\caption[]{Hubble diagram for the Type Ia SNe~\cite{C-T,SCP,SN1997}.
The residual magnitudes from the best-fit flat cosmology with a
cosmological constant and no photon loss
(the ``SCP accelerated universe'') are plotted.
The horizontal dashed line is the theoretical curve for the best-fit flat accelerated
cosmology with no axion-photon mixing
$(\Omega_m, \Omega_X, w_X)=(.28,.72,-1)$;
the solid line is the theoretical curve for the flat
cosmology with a choice of photon-axion mixing parameter
$(\Omega_m, \Omega_X, w_X, L_\mathrm{dec} H_0 /c) = (.3,.7,-1/3,1/3)$;
the lower dashed line shows the effect of turning off the photon-axion
coupling with the otherwise identical choice of parameters.
Finally, the upper dot-dot-dashed line is the remaining photon intensity
probability on a distance~$r_s(z) R_\circ$, with the probability scale given to the
right of the plot. The lower left plot emphasizes the very high $z$ behavior;
and
the lower
right plot enlarges the region near the farthest observed Type Ia SN.
}
\label{fig:Hubble}
\end{figure}

The data are displayed together in Figure~\ref{fig:Hubble}, together with
magnitude-redshift curves of various models including the CKT model.  The
data are displayed in terms of residual magnitudes with respect to the
SCP flat accelerated universe.
The CKT model differs only slightly from the SCP model for small $z$.
Note that
for $1<z<5$ the CKT magnitude dips below the SCP accelerated universe.
This fact
may allow the models to be distinguished if better high-$z$ data is obtained:
a precision of .05 in magnitude at $z\sim 1$ would allow
this distinction between the two scenarios to be made.
The deviation at $z>100$ should not be taken seriously, as the universal
magnetic field was likely insignificant before large scale structure formation
(but see for example \cite{Grasso-Rubinstein} for an alternative point of
view).

\begin{figure}[ht]
\centerline{\epsfxsize=15cm\epsfbox{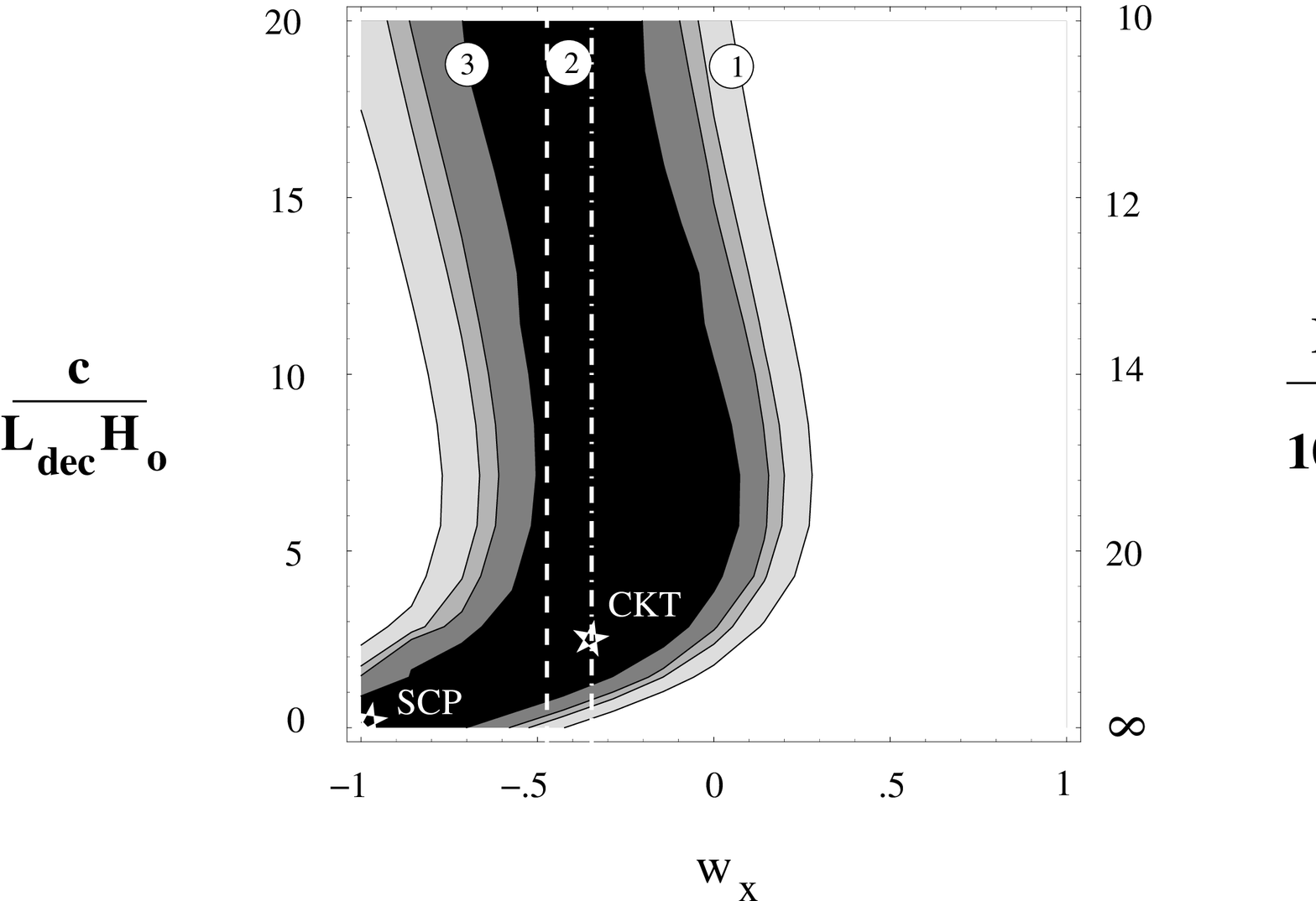}}
\caption[]{From black to white: $68\%$, $90\%$, $95\%$ and $99\%$
confidence regions in the
($w_X$\,--\,$L_\mathrm{dec}$) plane for a flat Universe
formed of matter with  $\Omega_{m}=.3$ and
an additional Dark Energy component with an equation of state $w_X$.
For a given amplitude of the intergalactic magnetic field and a given
length of coherence for it, the decay length is proportional to the square
of the axion coupling scale $M_L$ (the values reported
on the right correspond to $B=10^{-13}$ Tesla and $L_\mathrm{dom}=1$ Mpc).
The dashed and dot-dashed lines separate
regions of present acceleration (3),
present deceleration but future acceleration
(2) and eternal deceleration (1).
}
\label{fig:LdecWx}
\end{figure}

Figure~\ref{fig:LdecWx} demonstrates the maximum likelihood regions for
a flat universe with the CMB preferred value $\Omega_m=0.3$.
In this first statistical analysis, the effective theoretical
parameter space is two dimensional only, being spanned by the photon
decay length, $L_{\rm dec}$, and the equation of state of the Dark Energy
component, $w_X$.   Neglecting a photon-axion coupling
would have restricted ourselves to the line $L_{\rm dec}=\infty$ in the plot,
with the conclusion that at $99\%$ confidence level
the Universe is accelerating. Clearly a non-vanishing photon-axion coupling
opens up a valley that extends beyond $w_X<-1/3$ and thus allows
the SNe data to accomodate a Universe which does not accelerate,
either today or in the future.  Also note that even
with the enlarged parameter space due to the photon loss mechanism,
negative values for $w_X$ are strongly favored,
which is an independent result consistent with
the constraints coming from the CMB anisotropies and
structure formation.

\begin{figure}[ht]
\centerline{\epsfxsize=15cm\epsfbox{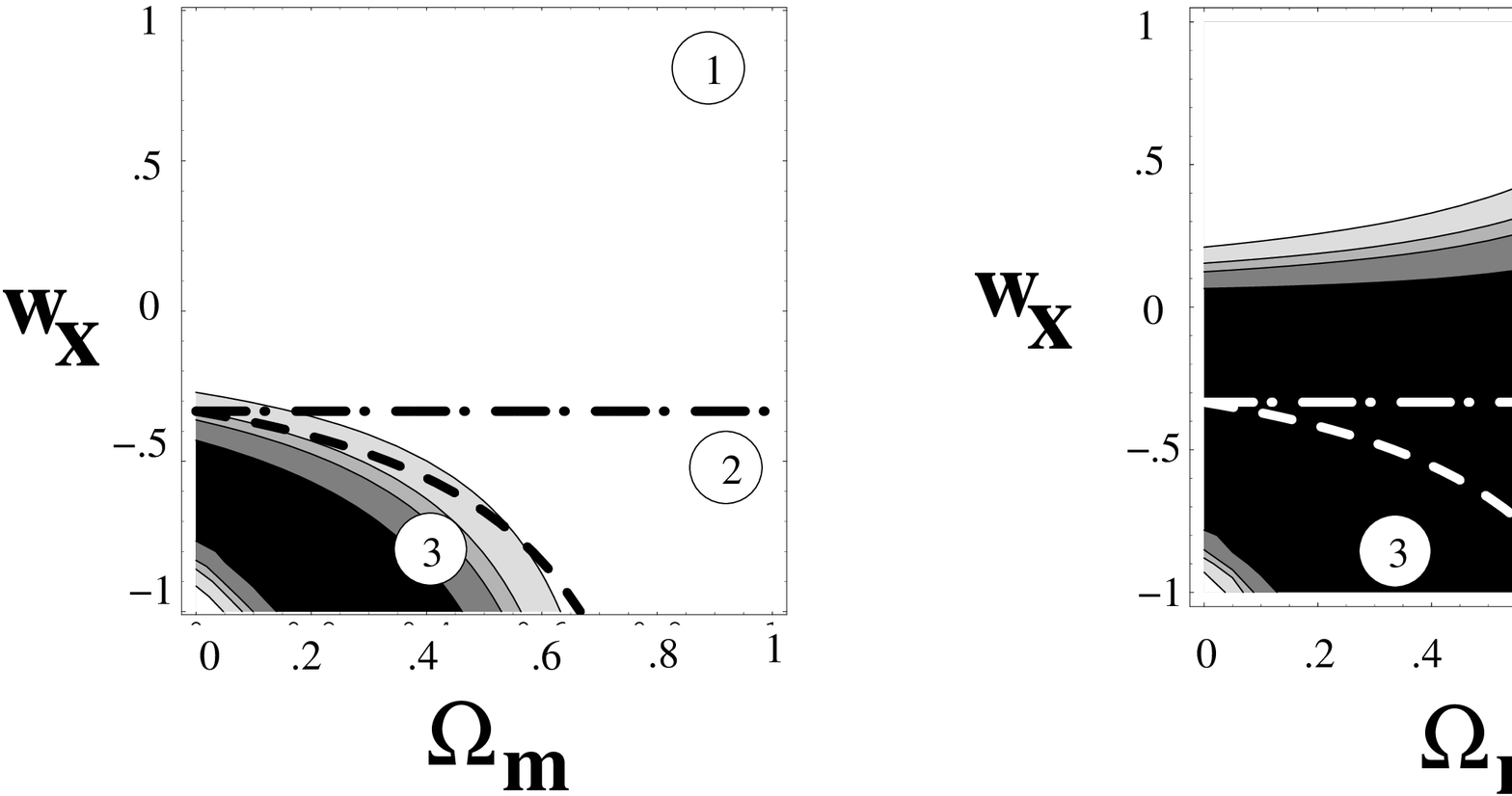}}
\caption[]{Best-fit, from black to white: $68\%$, $90\%$, $95\%$ and $99\%$
confidence regions in the
($\Omega_{m}$--\,$w_X$) plane
for a flat Universe.
The plot on the left is obtained in the standard model without mixing between
photons and axions; the plot on the right shows the modifications
due to the photon loss mechanism.  The dashed and dot-dashed lines separate
regions of present acceleration (3),
present deceleration but future acceleration
(2) and eternal deceleration (1).
}
\label{fig:OmWx}
\end{figure}

It is worth studying what can be learned from current SNe data
while relaxing some constraints coming from the other cosmological experiments.
Of course we expect a larger acceptable region of parameter space with
the additional photon loss parameter, but in particular it is interesting to
determine the most likely region in the enlarged parameter space.
First, let us vary the matter density ratio, $\Omega_m$, keeping
a flat Universe ($\Omega_X=1-\Omega_m$). The parameter space is now
three dimensional. To arrive at the 2D confidence levels plotted
in Figure~\ref{fig:OmWx}, for each point $(w_X,\Omega_m)$, we have minimized
$\chi^2$ with respect to $L_{\rm dec}$ (a minimization
over the range of values for $L_{\rm dec}$ allowed by the different
constraints on the photon-axion coupling and the intergalactic
magnetic field). The two figures illustrate the bias introduced in the
statistical analysis when removing the decay length as a free parameter.
Notice again that while the usual interpretation of the SNe data
(in the absence of photon loss) prefers a current period of acceleration
at the $90\%$ level, the plot on the right
shows that the CKT mechanism can easily accommodate a cosmological evolution
without acceleration.

\begin{figure}[ht]
\centerline{\epsfxsize=15cm\epsfbox{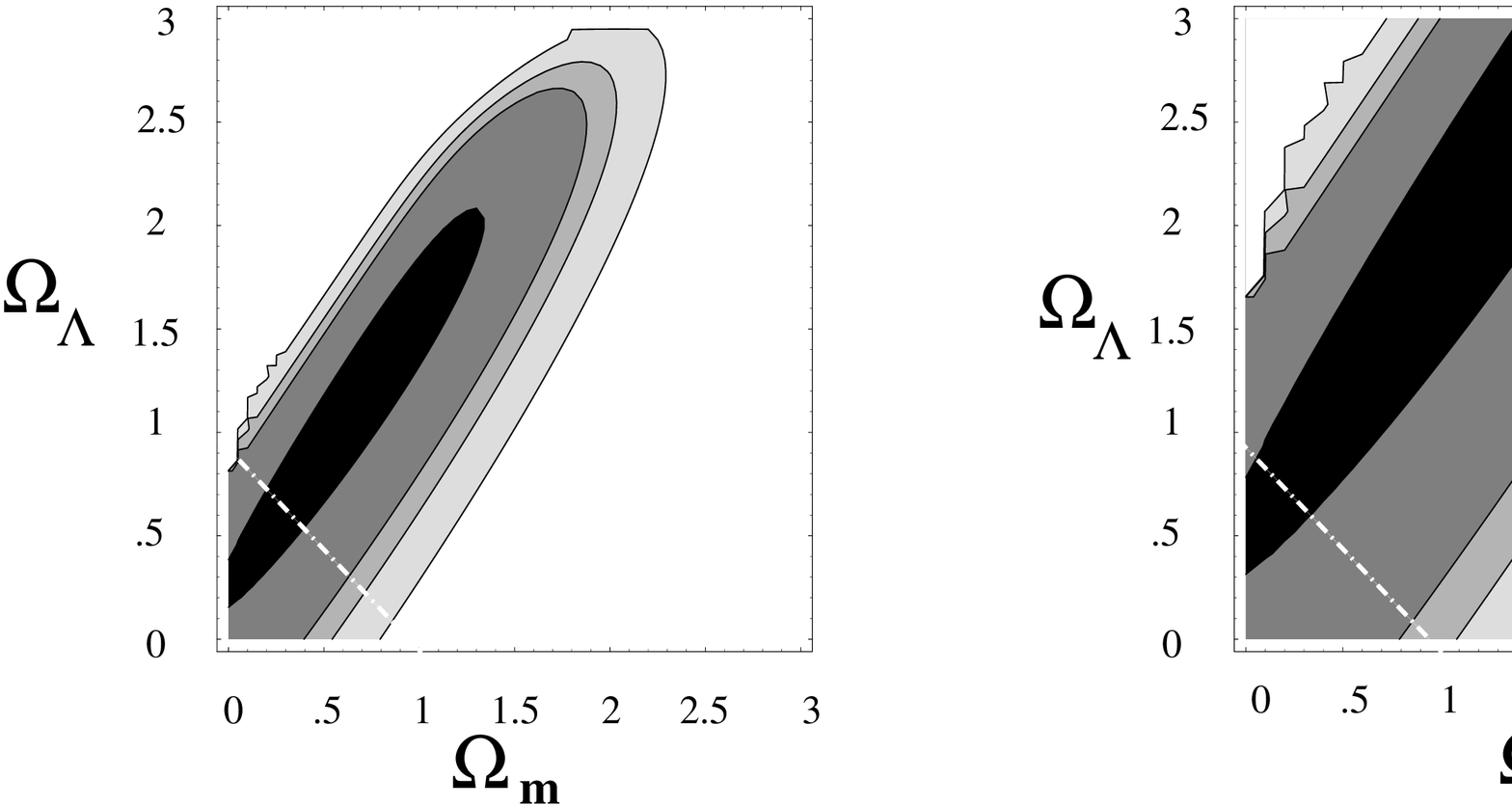}}
\caption[]{Best-fit, from black to white: $68\%$, $90\%$, $95\%$ and $99\%$
confidence regions in the
($\Omega_{m}$--\,$\Omega_{\Lambda}$) plane
for a      Universe
formed of matter and
a additional cosmological constant
The comparision of the two plots shows the influence of the
photon loss mechanism (on the left  the decay length of the photon is
set to infinity while on the right it is fit to its best $\chi^2$ value
at
each point).  The dot-dashed line locates flat universes.
}
\label{fig:OmOX}
\end{figure}

In the last fit reproduced in Figure~\ref{fig:OmOX}, we have assumed
that the Dark Energy is formed of a vacuum energy ($w_X=-1$) and we have
examined the bias in the determination of the remaining cosmological
parameters. Again, for each point $(\Omega_\Lambda,\Omega_m)$ in the plot
on the right of Figure~\ref{fig:OmOX}, $\chi^2$
is first minimized with respect to $L_{\rm dec}$. Of course, since
the dimming of SNe is now a combined effect of acceleration and the CKT
photon loss mechanism, the confidence level regions are enlarged
compared to the usual SNe data analysis.

\section{Conclusions}
\label{sec:conclusions}

We have presented an analysis of the SNe data taking into account
a photon loss mechanism as recently proposed by CKT.
This mechanism offers a new parameter, namely the photon decay length,
to accomodate the data. The main conclusion of our analysis,
as already announced by CKT, is that the dimming
of SNe at large redshift no longer requires
an accelerating expansion of the Universe, and
the photon-axion coupling opens up the window for
the equation of state of the Dark Energy towards less negative values.
We argue that the accelerating universe and the CKT models may be
distinguishable if better high-$z$ data becomes available.  In this way
future experiments like SNAP may reveal new physics beyond the
standard models of both particle physics and cosmology.
In the new allowed region of the parameter space, the SNe data could not only
confirm the existence of a Dark Energy component but may also
reveal the existence of a light pseudoscalar that has so far evaded
any direct particle physics detection.

\section*{Acknowledgments}
We are especially grateful to Csaba Cs\'aki, Nemanja Kaloper and
John Terning for sharing their work prior to publication, and for many
interesting discussions.
We are also happy to thank Tanmoy Bhattacharya, Guillaume Blanc, Salman Habib and
Hui Li for useful discussions.
C.G. thanks the hospitality of the T-8 group at Los Alamos, where this
work was performed.
J.E. is supported in part by the US Department of Energy
under contract W-7405-ENG-36.


\end{document}